\documentclass{article}

\usepackage{arxiv}

\usepackage[utf8]{inputenc} 
\usepackage[T1]{fontenc}    
\usepackage{hyperref}       
\usepackage{url}            
\usepackage{booktabs}       
\usepackage{amsfonts}       
\usepackage{nicefrac}       
\usepackage{microtype}      
\usepackage{lipsum}		
\usepackage{graphicx}
\usepackage{natbib}
\usepackage{doi}

\title{Mapping Tropical Forest Cover and Deforestation with Planet NICFI Satellite Images and Deep Learning in Mato Grosso State  (Brazil) from 2015 to 2021}

\author{ \href{https://orcid.org/0000-0002-9623-1182}{\includegraphics[scale=0.06]{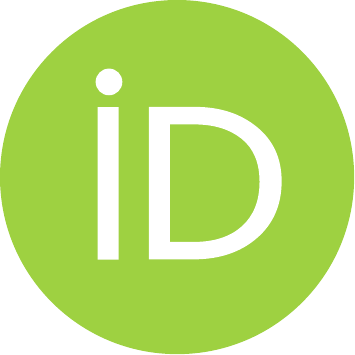}\hspace{1mm}Fabien H.~Wagner}\\
	Institute of the Environment and Sustainability,\\
	University of California, Los Angeles, CA 90095 USA;\\
	and	NASA-Jet Propulsion Laboratory\\
	California Institute of Technology, Pasadena, CA 91109, USA; \\ 
	and CTREES.org, Pasadena, United States;\\
	\texttt{wagner.h.fabien@gmail;fhwagner@ucla.edu} \\
 	\And
 	\href{https://orcid.org/0000-0002-7151-8697}{\includegraphics[scale=0.06]{orcid.pdf}\hspace{1mm}Ricardo Dalagnol} \\
	Institute of the Environment and Sustainability,\\
	University of California, Los Angeles, CA 90095 USA;\\
	and	NASA-Jet Propulsion Laboratory\\
	California Institute of Technology, Pasadena, CA 91109, USA ; \\ 
	and CTREES.org, Pasadena, United States;\\ 
 	\texttt{ricds@hotmail.com} \\
 	\And
 	\href{https://orcid.org/0000-0002-1052-5551}{\includegraphics[scale=0.06]{orcid.pdf}\hspace{1mm}Celso HL~Silva-Junior} \\
	Institute of the Environment and Sustainability,\\
	University of California, Los Angeles, CA 90095 USA;\\
	and	NASA-Jet Propulsion Laboratory\\
	California Institute of Technology, Pasadena, CA 91109, USA ; \\ 
	and CTREES.org, Pasadena, United States;\\ 
 	\texttt{csilva@ctrees.org} \\
 		\And
 	{\hspace{1mm}Griffin Carter} \\
 	Institute of the Environment and Sustainability,\\
	University of California, Los Angeles, CA 90095 USA;\\
	and CTREES.org, Pasadena, United States;\\
 	\texttt{gcarter@ctrees.org} \\
 		\And
 	{\hspace{1mm}Alison L~Ritz} \\
 	Virginia Polytechnic Institute and State University,\\
 	Interdisciplinary Graduate Education \\
 	Program in Remote Sensing,\\
 	Blacksburg, VA 24061, United States;\\
 	\texttt{alritz20@vt.edu} \\
 	 	\And
 	\href{https://orcid.org/0000-0002-1817-360X}{\includegraphics[scale=0.06]{orcid.pdf}\hspace{1mm}Mayumi CM~Hirye} \\
 	Quapá Lab\\
 	Faculty of Architecture and Urbanism\\
 	University of São Paulo---USP\\
 	São Paulo, SP, Brazil\\
 	\texttt{mayhirye@hotmail.com} \\
 	 	 	\And
 	  	\href{https://orcid.org/0000-0002-4221-1039}{\includegraphics[scale=0.06]{orcid.pdf}\hspace{1mm}Jean PHB~Ometto} \\
 	Earth Observation and Geoinformatics Division\\
 	National Institute for Space Research – INPE\\
 	São José dos Campos, SP, Brazil\\
 	\texttt{jean.ometto@inpe.br} \\
		\And
 	\href{https://orcid.org/0000-0001-8524-4917}{\includegraphics[scale=0.06]{orcid.pdf}\hspace{1mm}Sassan Saatchi} \\
	Institute of the Environment and Sustainability,\\
	University of California, Los Angeles, CA 90095 USA;\\
	and	NASA-Jet Propulsion Laboratory\\
	California Institute of Technology, Pasadena, CA 91109, USA \\ 
 	\texttt{sasan.s.saatchi@jpl.nasa.gov}; \\ 
	and CTREES.org, Pasadena, United States
}

\renewcommand{\shorttitle}{\textit{arXiv} Template}

\hypersetup{
pdftitle={A template for the arxiv style},
pdfsubject={q-bio.NC, q-bio.QM},
pdfauthor={David S.~Hippocampus, Elias D.~Striatum},
pdfkeywords={First keyword, Second keyword, More},
}

\begin{document}
\maketitle

\begin{abstract}
Monitoring changes in tree cover for rapid assessment of deforestation is considered the critical component of any climate mitigation policy for reducing carbon. Here, we map tropical tree cover and deforestation between 2015 and 2022 using 5 m spatial resolution Planet NICFI satellite images over the state of Mato Grosso (MT) in Brazil and a U-net deep learning model. The tree cover for the state was 556510.8 km$^2$ in 2015 (58.1 \% of the MT State) and was reduced to 141598.5 km$^2$ (14.8 \% of total area) at the end of 2021.  After reaching a minimum deforested area in December 2016 with 6632.05 km$^2$, the bi-annual deforestation area only showed a slight increase between December 2016 and December 2019. A year after, the areas of deforestation almost doubled from 9944.5 km$^2$ in December 2019 to 19817.8 km$^2$ in December 2021. The high-resolution data product showed relatively consistent agreement with the official deforestation map from Brazil (67.2\%) but deviated significantly from year of forest cover loss estimates from the Global Forest change (GFC) product, mainly due to large area of fire degradation observed in the GFC data. High-resolution imagery from Planet NICFI associated with deep learning technics can significantly improve mapping deforestation extent in tropics.
\end{abstract}


\section{Introduction}


Deforestation of tropical forests caused mainly by expanding forestry and agriculture is currently the second largest source of anthropogenic carbon emission, with 5.3$ \pm$ 2.4 GtCO$2$ emitted per year on the period 2001-2019, and still increasing  \cite{harris2021,friedlingstein2022,feng2022,IPCC2022,pendrill2019}. In the Brazilian Amazon, the deforestation rate of 2020 was the greatest of the decade so far\cite{silva2021}, and was recently overpass by the 2021 deforestation rate, with 13.032 km$^2$ \cite{Prodes2021}. Apart from carbon loss, $\sim$40 \% of this deforestation happens in old-growth tropical forests, causing also an increase in fragmentation \cite{montibeller2020}, and constitutes a direct threat to biodiversity \cite{harris2021}. Monitoring changes in tree cover for rapid assessment of deforestation is considered the key component of any climate mitigation policies for reducing emissions from deforestation and restoration of forests for biodiversity and carbon sequestration \cite{valeriano2004,mitchard2018}.

However, maps of tree cover and deforestation are still challenging datasets to produce. One of the most renown and accurate products of deforestation monitoring is the PRODES from the National Institute of Space Research of Brazil \cite{INPE2021}. Since 1988, the PRODES monitor clear-cut deforestation in the Amazon and produced annual deforestation rates for the Brazilian government. PRODES defines deforestation as the suppression of areas of old-growth forested physiognomies by anthropogenic actions, from intact forest cover to other lanf use \cite{Prodes2021}. It is produced based on the visual interpretation of satellite imagery and manual mapping of deforestation by trained specialists over the course of the PRODES-year, which starts at August 1$^{st}$ and ends on July 31$^{st}$ \cite{Prodes2021}. PRODES map is based on satellite dataset at 20 to 30 m of spatial resolution and 5 to 16 days of revisit rate (currently Landsat-8, SENTINEL-2 and CBERS-4/4A), and some pixels can have no observation due high cloud coverage during the course of one year. One of the limits of the product is that the manual sampling constrains the time of production of the map and limits the scale of the areas that can be considered as deforestation. For example PRODES does not map deforestation below 6.25 ha \cite{Prodes2021}. Another major product to monitor deforestation is the forest loss year dataset, a part of the Global Forest Change products (GFC), the first global map at 30 m of forest cover freely available globally from the University of Maryland \cite{Hansen850}. This product looks at the annual change in tree cover in relation to the tree cover of the year 2000 at 30 m spatial resolution globally. Forest loss is defined as the stand-replacement disturbance or the complete removal of tree cover canopy at the Landsat pixel scale. These two products, PRODES and GFC tree loss year product, are reference datasets to monitor deforestation; however, they are delivered annually and at 30 m of spatial resolution. Furthermore, Prodes is limited to deforested areas above 6.25 ha and GFC tree loss year include forest fires and forest degradation in the tree loss product, not only deforestation.

Recently, Planet satellite images over the tropics have been made available by the Norway’s International Climate and Forest Initiative (NICFI, \url{https://www.nicfi.no/}), to help save the world's tropical forests while improving the livelihoods of those who live off, in, and near the forests \cite{Planet2017}. The Planet NICFI images are multispectral satellite images containing red, green, blue, and near infrared bands at 4.78 m of spatial resolution for the  Normalized Analytic Basemaps. The temporal resolution of the NICFI images is currently one month, and they are a mosaic composite of the best daily acquisitions during the month. Consequently, Planet NICFI images are mostly cloud free, thus providing the best freely available multispectral dataset to monitor Land Use and Land Cover (LULC) changes in tropical regions.

However, while Planet NICFI provide the best spatial and temporal resolution to map tropical forest cover with optical data, the variation of reflectance values between the different Planet satellite sensors, between date and sometime within the same image is highly challenging. As state by the NICFI documentation, the absolute radiometric accuracy is not guarantee for the normalized surface reflectance basemaps \cite{Pandey2021}.  For the non-deep learning remote sensing analysis, where high quality reflectance values are needed, this is almost a no-go, unless several preprocessing steps are made, such as indices creations and bands normalisation \cite{cheng2020phenology,Francini2020}. However, for deep learning methods, the high quality or fidelity of reflectance values are not necessary. For example, to recognize  cars or plastic balloons in images, CNN models are even trained with images where the hue is artificially changed (during data augmentation) to impede the model to give too much importance to the color, which could be considered as overfitting in this case \cite{chollet2018deep}. Furthermore, colors (or values in the colour channels red, green, and blue) can be seen as one feature when we currently know that several features and multiple levels of abstraction are needed to reach state-of-the-art accuracy of classification \cite{lecun2015} as demonstrated by the success of CNNs in computer vision tasks. CNNs correspond exactly to what is needed here, to extract the maximum of information relevant to forest mapping using pixel context and not only pixel-based colors values with minimal pre-processing from the unequally calibrated images. Furthermore, deep learning has already been applied with success for tasks such as classification of cloud, shadows and land cover scenes in PlanetScope \cite{shendryk2019} and to map deforestation in Brazil with Landsat images \cite{matosak2022mapping,maretto2020sp}.

To segment evergreen forest tree cover in Planet images with deep learning, the U-net model was chosen \cite{Ronneberger2015}. This is a state-of-the-art model for the segmentation tasks, that returns the probability of the object to map per pixel and has shown excellent performance on tropical forest cover mapping in very high-resolution images \cite{Wagner2019}. Here, the tree cover model returns a binary value for each pixel, 0 or 1 for non-forest or forest, respectively, that constitute the tree cover mask. Then the deforestation date is deducted from the forest masks at each date. The model was tested in the Mato Grosso States in Brazil with the complete the Planet dataset from December 2015 to March 2022, corresponding to a total of 85056 images. The region located in the south of the Amazon Forest was chosen because it has overcome large amounts of deforestation the past. Between the years 2000 and 2013 there has been estimated  76,360 km2 of deforestation in Mato Grosso which corresponded to 32\% of the whole Brazilian Amazon estimates \cite{tyukavina2017}. Furthermore,  Mato Grosso state also contains some of the main current deforestation fronts of the region know as the arc of deforestation of the Brazilian Amazon forest.

In this work is presented (i) the product of evergreen tree cover and deforestation obtained with Planet images and deep learning, (ii) the validation of the product with independent tree cover mask based on Planet data and also from an independent LiDAR dataset, and (iii) the comparison with PRODES deforestation and GFC tree loss year loss products.

\section{Materials and Methods}

  \begin{figure}[ht]
 \centering
 \includegraphics[width=0.7\linewidth]{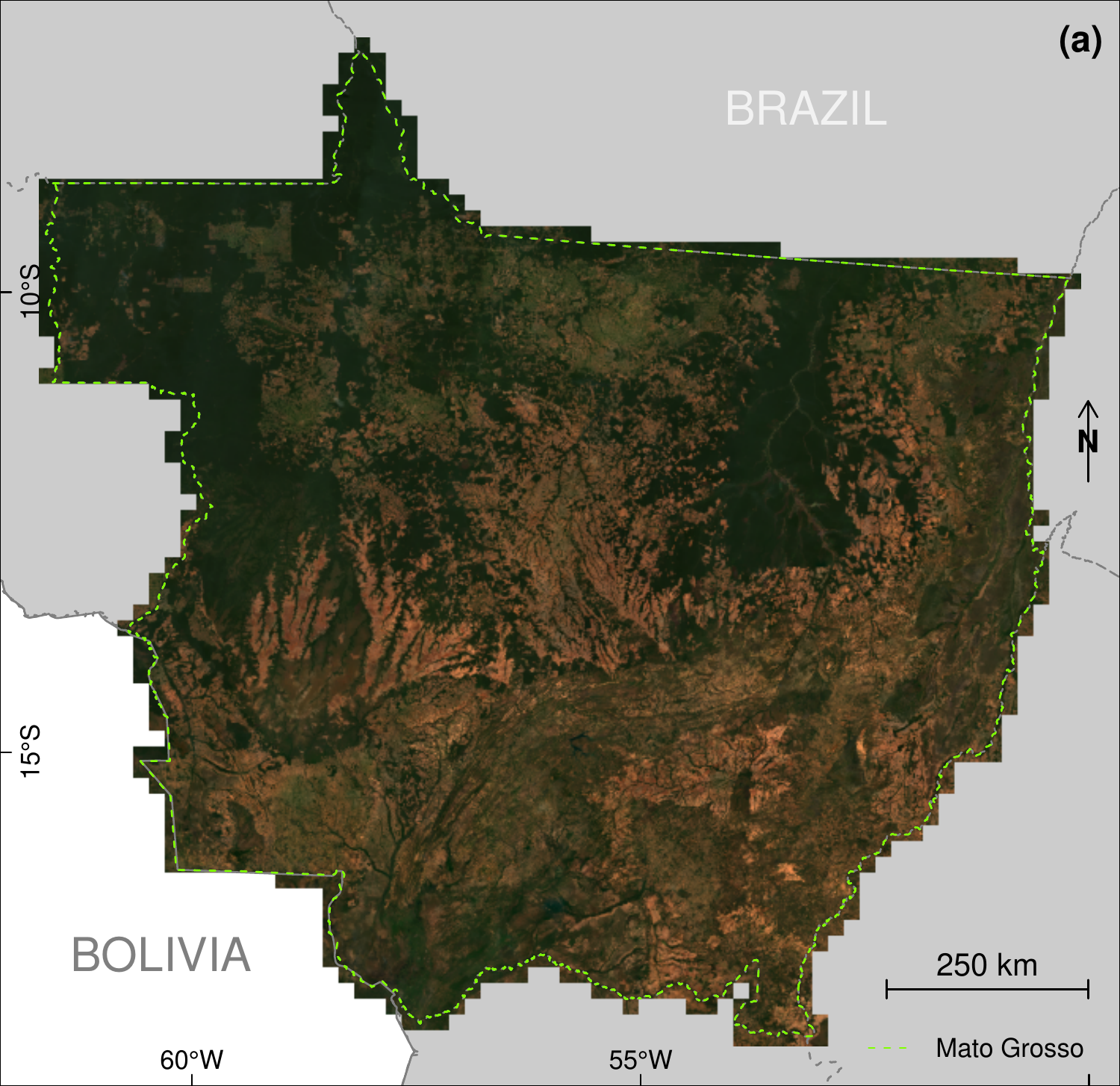}
  \caption{Planet NICFI mosaic of September 2021 constituted of the 2658 Planet image tiles covering the Mato Grosso State - Brazil.}
  \label{Fig1}
  \end{figure}
  
\subsection{Planet satellite images of Mato Grosso - Brazil}\label{satimg}

There were 2658 Planet tiles of $\sim$ 20 $\times$ 20 km at $\sim$ 4.78 m spatial resolution covering the entire Mato Grosso State of Brazil, Fig \ref{Fig1}. This study region cover represents $\sim$ 1 millions of km$^2$. Images were downloaded trough the Planet API \url{https://api.planet.com/basemaps/v1/mosaics} and with the PlanetNICFI R package \cite{Mouselimis2022} for all the 32 available dates \cite{Planet2017} at the time of the study. The complete dataset contains 85056 planet images for the following dates, bi-annually: 2015-12-01, 2016-06-01, 2016-12-01,  2017-06-01, 2017-12-01, 2018-06-01, 2018-12-01, 2019-06-01, 2019-12-01, 2020-06-01; and, monthly from  2020-09-01 to 2022-06-01. All bands in raw image digital numbers (12 bits),Red (0.650-0.682  $\mu$m), Green (0.547-0.585 $\mu$m), Blue (0.464 -0.517 $\mu$m) and the NIR bands (0.846-0.888 $\mu$m) \cite{planet2021}, were, first, truncated to the range 0--2540 for the RGB bands and scaled between 0 and 2540 for the NIR bands (i.e. divided by 3.937). Second, the 4 bands were scaled to 0--255 (8 bits) by dividing by 10 and then the Red-Green-Blue-NIR (RGBNIR) composite was built. The forest reflectance values are low in the RBG bands ($<$ 500) and the specific scaling of these bands was made optimize the range of values of the forest reflectance in 8 bits. The scaling of NIR is only a min-max (0-10000) scaling as forest reflectance values are not low in this band. No atmospheric correction was performed. A second image was generated from the composite adding a mirroring border of 128 pixels on each side for the deep learning prediction in order to remove border effects.

\subsection{LiDAR tree cover dataset}
To validate our tree cover map, we compared our results to the LiDAR height data that were acquired in the scope of the Amazon Biomass Estimate project (EBA/INPE) which covers mainly Amazonian forests. The complete dataset is constituted of discrete-return LiDAR data with 610 flightlines of 15 km long by 0.5 km wide distributed randomly over the amazon region ($\sim$ 787 ha each; total area of 480,000 ha). Among this dataset, we selected 141 flightlines that intersected entirely with the Planet images over Mato Grosso. The LiDAR dataset was acquired during 2016 using the Trimble HARRIER 68i laser scanning system onboard an airplane at an average flight altitude of 600 m. Multiple LiDAR returns were recorded with a minimum point density of 4 points per m$^2$. The horizontal and vertical accuracy ranged from 0.035 m to 0.185 m and from 0.07 m to 0.33 m, respectively. It is currently the largest available non-continuous dataset of high-resolution airborne LiDAR over Amazon forests The LiDAR point clouds were processed into digital terrain models (DTM) and canopy height models (CHM) with 1 $\times$ 1 m cell size, as described by \cite{dalagnol2019quantifying,dalagnol2021} and the median of the data was aggregated at the resolution of the planet data.

\subsection{Deforestation history datasets}
To test if our tree cover and deforestation products were  consistent with independent datasets over the region, we compared our results to two existing products. The first product is PRODES from INPE - the Brazilian National Institute for Space Research \cite{INPE2021}. The PRODES project monitors clear-cut deforestation in the Brazilian Legal Amazon based on satellite images of $\sim$20 to 30 m of spatial resolution (Landsat-8, CBERS-4 and similar imagery) and delivers official annual rates of forest loss for Brazil (1988-2021). The publicly available annual maps are all corrected/edited manually by experts and contain polygons with the following labels: forest, non-forest, deforestation of the year, previous deforestation, clouds, and water. The minimum area considered for mapping deforestation is 6.25 ha.
The second product is the forest loss year component of the Global Forest Change map of the University of Maryland at 30 m spatial resolution based on Landsat data \cite{Hansen850}, \url{https://developers.google.com/earth-engine/datasets/catalog/UMD_hansen_global_forest_change_2021_v1_9}. Forest Cover Loss is defined as a stand-replacement disturbance, or a change from a forest to non-forest state, during the period from 2000 to 2021.  
Both datasets were compared to our product (i) by computing the total area intersecting with our product at each time period, and (ii) by computing the percentage of intersections of GFC tree loss year or PRODES with our product by tiles weighted by the number of observations in GFC or PRODES.



\subsection{Neural Network Architecture}

\begin{figure}[ht]
\centering
\includegraphics[width=1\linewidth]{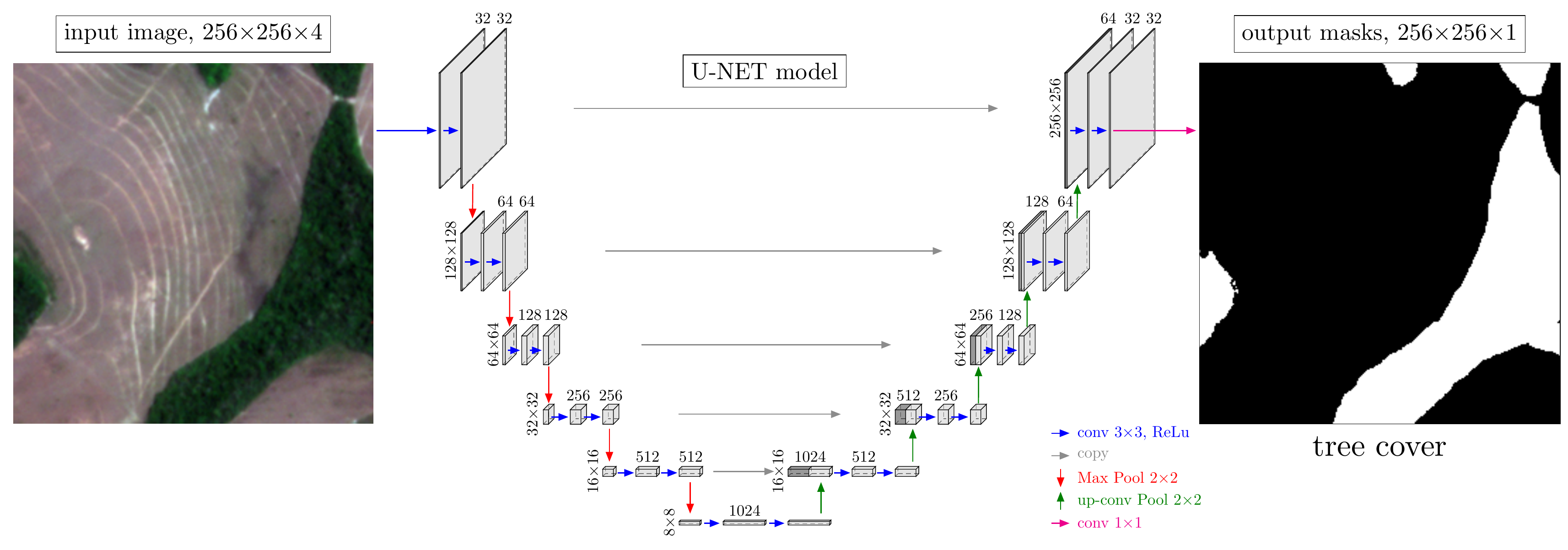}
 \caption{Architecture of the U-net model used for evergreen tropical forest mapping in Planet data.}
 \label{FigUnet} 
 \end{figure}

 The segmentation of evergreen forest cover in the Planet images of Mato Grosso was made with a classical U-net model \cite{Ronneberger2015}, Fig. \ref{FigUnet}. This model provides state of the art performance for segmentation of remote sensing images and have been already used to mapping forest cover in the tropics \cite{Huang2018,Wagner2019,wagner2020m}. More specifically the U-net model return the probability of presence of evergreen tropical forest in each pixel of a given input image. In our case, the model inputs are 4 bands RGB-NIR images made up of 256 $\times$ 256 pixels at $\sim$4.77 m of spatial resolution and the output a mask of one band and 256 $\times$ 256 pixels containing 1 (forest cover, pixel probability $>=$0.5) or 0 (non-forest, probability $>=$0.5). The model has a total of 34,614,657 parameters, of which 34,600,641 are trainable. The model was coded in R language \cite{CoreTeam2016} with RStudio interface to Keras and TensorFlow 2.8 \cite{chollet2015keras,AllaireChollet,allaireTang,AbadiAgarwalBarhamEtAl2015}.


\subsection{Training}
To produce the training sample of evergreen forest cover for the U-net model, 10 Planet images where randomly selected over Mato Grosso for each monthly Planet mosaic dates from September 2020 to September 2021 included, to constitute a dataset of 130 images. Among this dataset, 75 image were then selected after removing images with a high cloud cover or large variations in illumination. Then for the selected images, we ran the k-textures model \cite{Wagner2022} to perform image self-segmentation with 8 classes. We kept only 23 images where the forest segmentation was deemed accurate according to the visual interpretation of the forest and non-forest areas. Small errors of the forest mask were manually corrected. At this step, the masks had only two classes: forest (1), and non-forest (0) which contained agriculture, urban area, water surface and bare ground.
From this dataset, we trained a first U-net model \cite{Ronneberger2015} and applied predict tree cover over the original 130 images. Then, we removed the 23 images used in the first training sample, and selected a second set of 21 images with atmospheric conditions such as cloud and haze. The forest masks were selected or manually corrected to keep forests with no haze or very thin transparent cloud, while removing all clouds and heavy haze. Here, we want the model to classify all cloud and haze as non-forest when forests are difficult to recognize even by the human eye, keeping only the forests that a human can easily say it is forest cover. 
From the first and the second samples, we constitute a second training sample of 6663 images patches of 256 x 256 pixels and their associated labelled masks for the final U-net model. 6187 were containing forest and background and 476 only background. 80\% (5331) were used for training and 20\% (1332) for validation.

Each image patch goes through a data augmentation process that consists in random vertical and horizontal flips. No additional data augmentation was necessary because of the natural data augmentation provided by different atmospheric conditions and illumination due the different dates of the sampling images \cite{Wagner2021}. After data augmentation, the images were then fed to the U-net model.

During network training, we used a standard stochastic gradient descent optimization. The loss function was designed as a~sum of two terms: binary cross-entropy and Dice coefficient-related loss of the three predicted masks \cite{Dice1945,AllaireChollet,chollet2015keras} and finally the optimizer Adam \cite{Kingma2014} with a learning rate of 0.0001 was used. We used the accuracy (i.e. the frequency with which the prediction matches the observed value) as the metrics to assess the model performance. The network was trained for 20000 epochs with a batch size of 256 images and the model with the best weighted accuracy was kept for prediction (epoch 17058 and training/validation accuracy of 99.40 and 97.70\% respectively and training/validation loss of 0.0088 and 0.0574 respectively,). The training of the models took approximately 12.5 hours using a Nvidia RTX2080 Graphics Processing Unit (GPU) with an 8 GB memory. 


  
 \subsubsection{Prediction}
 
 For prediction, the Planet tiles of 4096 $\times$ 4096 pixels a border of 64 neighboring pixels were added on each side with mirroring image. This border method was used to avoid border artefacts during prediction, a known problem for the U-net algorithm \cite{Ronneberger2015}. Then the prediction was made on the entire image. To~belong to the evergreen forest class, the pixel prediction value must be greater than or equal to 0.5. The prediction for one Planet tile took approximately 7 seconds using a Nvidia RTX3090 GPU.

\subsubsection{Cloud temporal filter}
Even with the high temporal resolution of Planet satellite image acquisition (daily), the monthly mosaic  product can still contains some clouds. To filter for these clouds, we designed a simple temporal filter: to be classified as an evergreen forested pixel, a pixel at a given date must have been detected as evergreen forest cover at least half of the observations on the period defined by the closest 3 dates before and after its date. For the borders of the time series, we used the all the available dates in the $\pm$ 3 months around the date to filter.  

\subsubsection{Tree cover and deforestation masks}
At this step, we considered that the mask images filtered by cloud are almost cloud free and we give only a maximum of two chances to have a forest that have been misclassified as non-forest on the 2015-2021 period due to cloud. For example, if on the period of the cloud filter ($\pm$ 3 months), the forested pixel has never been observed due to clouds.

Then, we process the time series from the present to the past to reconstruct tree cover map per pixel. we keep only biannual time series as this point. 
For each pixel at each date, we keep or correct the value of the pixel with rules described in the following text and we attribute a value of trust of the classification, trust = 1 for "confirmed", the pixel is tree cover or deforestation, trust = 2 for "very likely", when a pixel is tree cover on all the data, non-tree cover on only 1 or 2 dates and with this dates being not the first and the last, note that the non-forested pixel values are all set to tree cover in this case. And finally trust = 3 for "unconfirmed". This value is only for the deforestation of the last available date, to be confirmed with next available date. Deforestation is defined by a pixel classified as non forest at the date N. The data are processed starting from the last date to the first date and the rules are applied at each date. Deforestation is also confirmed during the process, when a pixel at N-1 is classified as deforested and still deforested at N, is confirmed as deforestation and its trust value become 1. The result of this steps are corrected tree cover and deforestation masks for all the dates and their associated values of classification confidence. 

\section{Results}

\subsection{Tree cover validation with independent validation sample}
The tree cover algorithm presented a high level of segmentation accuracy, with an overall accuracy of 98.11\% and an F1-Score of 0.982 (precision = 0.975, recall = 0.988) on the 1332 images patches (256x256 pixels) of the validation sample.


\subsection{Tree cover validation with airborne LiDAR data}

\begin{figure}[ht]
 \centering
 \includegraphics[width=0.5\linewidth]{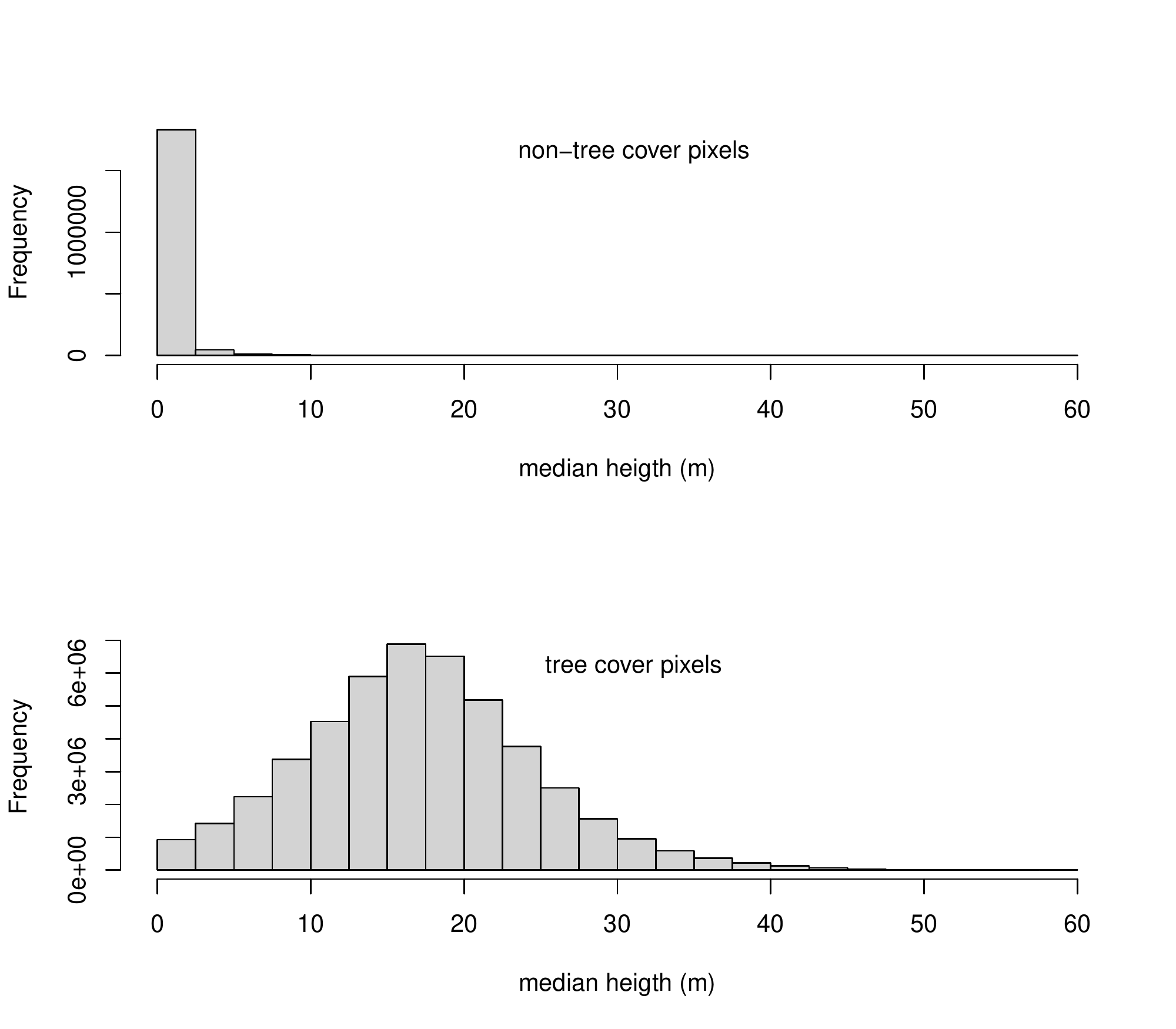}
  \caption{Distribution of median height in the Planet pixels that intersects  (with full overlap) with 141 EBA LiDAR datasets randomly distributed over Mato Grosso, for planet pixels classified by the U-net as non-tree cover (a) and tree cover (b), respectively. }
  \label{Fig2}
  \end{figure}

The validation of our tree cover product was made by comparing the median height measured by LiDAR at 1 m spatial resolution inside the planet pixels overlapping entirely with the 141 EBA LiDAR footprints above Mato Grosso. it represented 47267739 pixels for tree cover and 1912151 pixels for non-tree cover. In the non-tree cover pixels classified by the U-net, very few pixels have a median high above 2.5 meters, Fig. \ref{Fig2}a.

We found that 94 \% of the non-treecover pixels had a value below 0.55 m,  1 \% in the values ranging between 0.55 m and 2.07 m, 3 \% in the range between 2.07 m and 4.45 m and only 2\% above 4.45 m. For the tree cover pixels, the height was a distributed in a Gaussian shape around the median height of 16.9 m, Fig. \ref{Fig2}b, and 5 \% of the tree cover pixels has median height below 5.02 m. overall, based on the LiDAR data, it is observed that our tree cover model segment trees with a very high accuracy, that is, 98\% non forest pixels have height below 5 m, and around 95 \% of the tree cover pixels have height above 5 m.


\subsection{Tree cover and deforestation of Mato Grosso on the period 2015-12-01 to 2021-12-01}

For the Mato Grosso State, a tree cover of 556510.8 km$^2$ was found for the year 2015, which represent 58.1 \% of the State area, Table \ref{Tab1}.  Between December 2015 and December 2021, this area was diminished of $\sim$ 141598.5 km$^2$ (14.8 \% of total area). The biannual deforestation data did not show a clear seasonal preference for deforestation. After reaching the minimum value of deforestation area in December 2016 with 6632.05 km$^2$, the biannual deforestation area showed a tendency to increase between December 2016 and December 2019 and a sharp increase was observed thereafter, with the area of December 2021 (19817.8 km$^2$) been almost the double of the area of December 2019 (9944.5 km$^2$). Note that new forests growing after 2015 are not accounted in these numbers. Furthermore, even with the 6-months cloud filter some clouds can remains in the last date, and results for 2021-12-01 should be taken with caution.



\begin{table}[ht]
\centering
\begin{tabular}{lrr}
  \hline
classes & Area (km²) & Area (\%) \\ 
  \hline
  2015 total tree cover  & 556510.76 & 58.08 \\ 
  2015 non-tree cover & 401689.23 & 41.92 \\ 
  2015 remaining tree cover & 414912.25 & 43.30 \\ 
  deforestation 2016-06 & 18763.20 & 1.96 \\ 
  deforestation 2016-12 & 6632.05 & 0.69 \\ 
  deforestation 2017-06 & 8511.05 & 0.89 \\ 
  deforestation 2017-12 & 8037.00 & 0.84 \\ 
  deforestation 2018-06 & 8066.52 & 0.84 \\ 
  deforestation 2018-12 & 9016.48 & 0.94 \\ 
  deforestation 2019-06 & 8812.72 & 0.92 \\ 
  deforestation 2019-12 & 9944.50 & 1.04 \\ 
  deforestation 2020-06 & 9998.04 & 1.04 \\ 
  deforestation 2020-12 & 15297.82 & 1.60 \\ 
  deforestation 2021-06 & 18701.35 & 1.95 \\ 
  deforestation 2021-12 & 19817.78 & 2.07 \\ 
   \hline
\end{tabular}
 \caption{Tree cover and its change due to deforestation on the period from 2015-12-01 to 2021-12-01 obtained with Planet data and the U-net tree cover model.}
\label{Tab1}
\end{table}


  \begin{figure}[ht]
 \centering
 \includegraphics[width=1\linewidth]{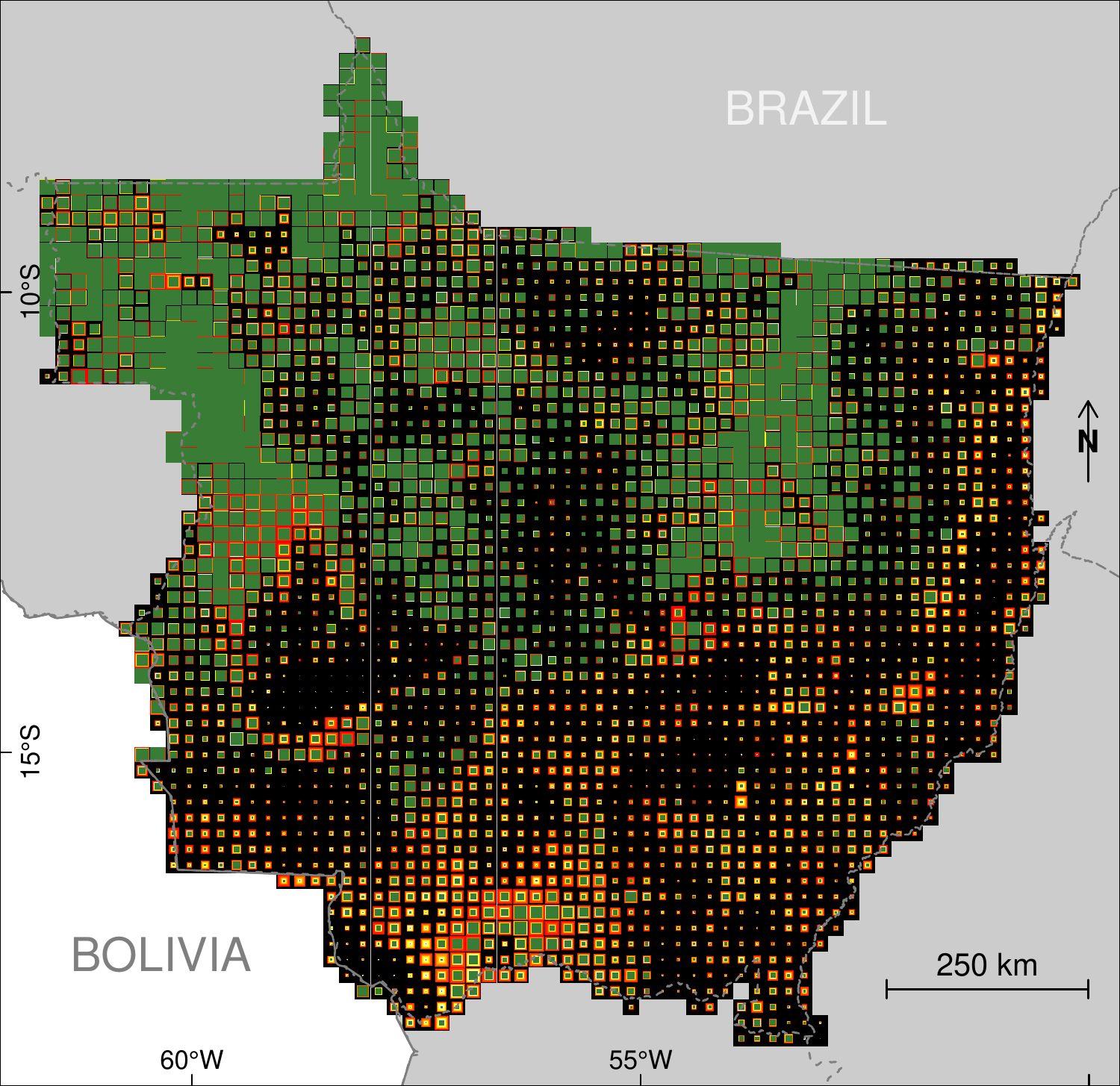}
  \caption{Tree cover and tree cover loss per year for the 2658 Planet tiles covering the Mato Grosso State - Brazil, with area represented with the real size found in each tiles. Remaining tree cover of 2015 is in green, and the yellow to red color scale represent the deforestation with the red being the more recent.} 
  \label{Fig3}
  \end{figure}

Only 52 (1.96 \%) of the 2658 Planet tiles of Mato Grosso States have less than 1 km$^2$ of tree cover change on the period between 2015 and 2021, Fig. \ref{Fig3}. The mean deforested area was 57.2 km$^2$ per tiles (tile area is 382.9019 km$^2$). 455 tiles (17.1 \%) shown deforested area above 100 km$^2$. Some areas with current intensive tree cover change are clustered such as areas located in the south of Mato Grosso, and near the largest forest remnants.  

 \begin{figure}[ht]
 \centering
 \includegraphics[width=0.9\linewidth]{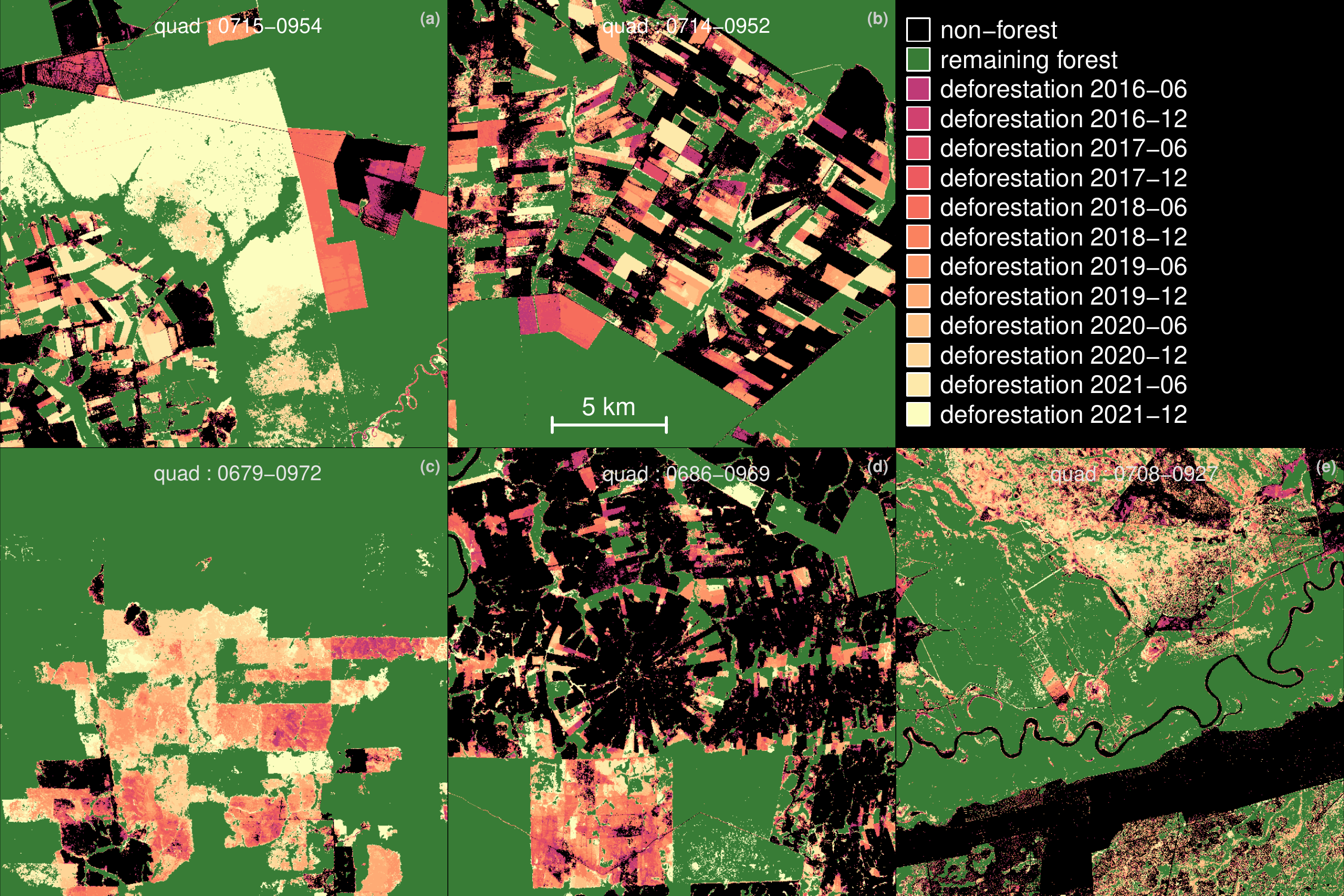}
  \caption{Example of our deforestation product for five Planet quad on the period 2015-12-01 to 2021-12-01.} 
 \label{Fig9}
 \end{figure}

Large scale deforestation still appeared in 2021, for example, in the Xingu region, Fig. \ref{Fig9}a. Note that roads built previously for logging appeared in the largest patch. Depending on the size of the rivers inside forest, it can take some times to appear clearly in the images due to satellite view angle, and small rivers can be mapped as deforestation in the first years, such as the river in the bottom right. Gradually advancing smaller scale deforestation can also be observed in the Xingu region, Fig. \ref{Fig9}b, accompanied by new roads inside forests. Large scale deforestation is visible on the recent deforestation front between the Mato Grosso and Amazonas states, Fig. \ref{Fig9}c. Some isolated dots are logging decks or roads and indicate where the deforestation will likely appear in a near future. An unusual circular pattern of deforestation for Mato Grosso shows still active deforestation in 2021, Fig. \ref{Fig9}d. It is located in the deforestation front between the Mato Grosso and Amazonas states. Some pattern of tree cover change can be natural or not, for example, in the South Central region of Mato Grosso, Fig. \ref{Fig9}e. Natural changes in river border were mapped as deforestation and this region also present new roads, and large areas of tree cover change. For this region, the large deforestation could be linked to the change from biannual to monthly time series and inclusion as forest when they are deciduous and now observed as deforestation. Note that not all this deforestation is necessary illegal. However, for most of these images small new roads or logging decks located inside the forest can be observed, indicated that logging is active and planned to continue.


\subsection{Similarities and discrepancies of our product with Prodes deforestation and GFC tree loss year data in the Amazon region of Mato Grosso}

In this section, we detailed which deforested pixels detected by Prodes or GFC tree loss year are found or not by our product for the Legal Amazon region of the Mato Grosso State.



\begin{table}[ht]
\centering
\begin{tabular}{rrrr}
  \hline
 &  & Deforestation & \\ 
 & our product (km$^2$) & Prodes (km$^2$) & GFC tree loss year (km$^2$) \\ 
  \hline
2016 & 7762.19 & 1370.87 & 5192.88 \\ 
  2017 & 5453.75 & 1327.95 & 7638.69 \\ 
  2018 & 5959.05 & 1407.31 & 3950.39 \\ 
  2019 & 6271.88 & 1848.54 & 3666.78 \\ 
  2020 & 7638.07 & 1817.41 & 5636.31 \\ 
  2021 & 13930.53 & 1860.23 & 3987.56 \\ 
   \hline
\end{tabular}
\caption{Annual tree cover change in the Amazonian region of Mato Grosso in km$^2$ on the period from 2016 to 2021-12-01 obtained with our product, Prodes data and GFC tree loss year data.}
\label{tab1.2}
\end{table}

For the tree cover change in the Amazon for region of Mato Grosso, Table \ref{tab1.2}, we found more tree cover changes that the Prodes and GFC tree loss year product. We found between 3.39 to 7.49 times more deforestation than Prodes depending on the date and the largest discrepancies were found on the last date of our time series. We found between 0.71 and 3.49 times the deforestation from GFC tree loss year product, with the last date having the largest differences. Note that the spatial resolution of Planet is 4.77 m while the spatial resolution of Prodes data and GFC tree loss is 30 m, and that even with the 6-months cloud filter some clouds can remains for the last date of our product, so results for 2021-12-01 should be taken with caution.


Because of the $\pm$ 3 dates aggregation for cloud filtering, our deforestation data start to be consistent in time with PRODES only when monthly Planet data are available. On the period 2020-06 to 2021-12, our model intersects with 72.5\% of the Prodes data for the year 2021, Table \ref{tab2}. On the period 2019-06 to 2021-06, our model intersects with 64.4\% of the Prodes data for the year 2020.  On the period 2018-06 to 2020-06, our model intersects with 60.7\% of the Prodes data for the year 2019. The largest class outside the intersection of dates is always forest cover, meaning that we have classified as tree cover a pixel that was classified as deforested by Prodes. This can be errors from our model or small patches of forests that are present in the 30 m pixels classified as deforestation by Prodes.

\begin{table}[ht]
\centering
\resizebox{\textwidth}{!}{
\begin{tabular}{lrrrrrrrr}
  \hline
   & &  &   &  PRODES &    &   &    &    \\ 
 & non-forest & forest & deforestation  & deforestation  & deforestation   & deforestation  & deforestation   &   deforestation \\ 
  &  &  &  2016-12 &  2017-12 &   2018-12 &  2019-12 &   2020-12 &   2021-12 \\ 
  \hline
 OURS  & &  &   &   &    &   &    &    \\ 
  \hline
non-forest & 67.30 & 1.30 & 52.60 & 22.20 & 9.30 & 6.00 & 5.00 & 3.50 \\ 
  forest & 19.50 & 95.30 & 8.60 & 9.90 & 13.60 & 11.60 & 17.90 & 17.20 \\ 
   \hline
    deforestation  & &  &   &   &    &   &    &    \\ 
  2016-06 & 2.10 & 0.20 & 8.00 & 15.50 & 3.00 & 1.40 & 1.10 & 0.60 \\ 
   2016-12 & 0.70 & 0.10 & 1.90 & 1.70 & 0.50 & 0.40 & 0.30 & 0.20 \\ 
   2017-06 & 0.90 & 0.10 & 4.40 & 13.60 & 4.20 & 1.40 & 0.80 & 0.50 \\ 
   2017-12 & 0.80 & 0.10 & 3.30 & 9.80 & 13.40 & 2.40 & 1.00 & 0.50 \\ 
   2018-06 & 0.90 & 0.10 & 3.10 & 5.30 & 16.80 & 4.20 & 1.50 & 0.60 \\ 
   2018-12 & 0.90 & 0.10 & 3.80 & 4.60 & 11.40 & 18.70 & 2.70 & 0.90 \\ 
   2019-06 & 0.80 & 0.10 & 3.20 & 3.20 & 7.70 & 18.60 & 7.20 & 1.40 \\ 
   2019-12 & 0.90 & 0.20 & 2.70 & 3.30 & 5.00 & 12.50 & 19.80 & 2.10 \\ 
   2020-06 & 0.80 & 0.20 & 1.90 & 2.70 & 3.20 & 6.70 & 16.30 & 3.60 \\ 
   2020-12 & 1.20 & 0.40 & 2.70 & 3.40 & 5.50 & 7.40 & 11.80 & 20.20 \\ 
   2021-06 & 1.60 & 0.50 & 2.10 & 2.70 & 3.90 & 5.50 & 9.30 & 34.00 \\ 
   2021-12 & 1.50 & 1.40 & 1.70 & 2.10 & 2.70 & 3.30 & 5.50 & 14.70 \\ 
   \hline
\end{tabular}
}
\caption{Intersection in percentage of annual Prodes data and our bi-annual deforestation product.}
\label{tab2}
\end{table}

Regarding the GFC tree loss year data, on the period 2020-06 to 2021-12, our model intersects with 57.9\% of the GFC tree loss year data, Table \ref{tab2}; on the period 2019-06 to 2021-06  with 33.9\% of the GFC tree loss data for the year; and, on the period 2018-06 to 2020-06  with 36.7\% of the GFC tree loss data. The largest class is always forest cover, meaning that we have classified as tree cover a pixel that was classified as tree loss in GFC data. These are errors from our model and are also expected because GFC tree loss year consider a change inside the Landsat pixels, but this does not always represent a clear cut of the forest in the pixel, only a significant change in tree cover.




\begin{table}[ht]
\centering
\resizebox{\textwidth}{!}{
\begin{tabular}{lrrrrrrrr}
 \hline
   & &  &   &  GFC tree loss year &    &   &    &    \\ 
 & non forest & forest 2000 & deforestation  & deforestation  & deforestation   & deforestation  & deforestation   &   deforestation \\ 
  &  &  &  2016-12 &  2017-12 &   2018-12 &  2019-12 &   2020-12 &   2021-12 \\ 
  \hline
 OURS  & &  &   &   &    &   &    &    \\ 
  \hline
non forest & 26.90 & 72.90 & 26.30 & 7.80 & 5.40 & 4.90 & 3.90 & 5.30 \\ 
  forest & 67.40 & 13.10 & 42.30 & 50.30 & 38.70 & 36.50 & 56.40 & 31.60 \\ 
    \hline
 deforestation  & &  &   &   &    &   &    &    \\ 
  2016-06 & 0.80 & 2.70 & 5.40 & 5.30 & 2.20 & 1.10 & 0.60 & 0.90 \\ 
  2016-12 & 0.30 & 0.70 & 1.20 & 0.70 & 0.30 & 0.60 & 0.30 & 0.40 \\ 
  2017-06 & 0.30 & 1.20 & 2.70 & 4.60 & 2.80 & 0.90 & 0.40 & 0.50 \\ 
  2017-12 & 0.30 & 1.00 & 2.20 & 5.00 & 5.30 & 1.40 & 0.40 & 0.60 \\ 
  2018-06 & 0.30 & 1.00 & 1.90 & 3.10 & 9.00 & 1.90 & 0.50 & 0.70 \\ 
  2018-12 & 0.30 & 1.20 & 2.50 & 3.10 & 9.20 & 8.90 & 0.90 & 0.60 \\ 
  2019-06 & 0.30 & 1.00 & 2.20 & 2.60 & 5.90 & 11.40 & 1.50 & 0.60 \\ 
  2019-12 & 0.30 & 1.00 & 2.30 & 2.50 & 4.50 & 9.40 & 6.00 & 0.80 \\ 
  2020-06 & 0.40 & 0.70 & 1.70 & 2.20 & 2.90 & 5.10 & 7.00 & 1.00 \\ 
  2020-12 & 0.50 & 1.10 & 2.80 & 4.30 & 5.10 & 6.00 & 13.50 & 6.20 \\ 
  2021-06 & 0.80 & 1.30 & 2.60 & 3.90 & 4.20 & 6.00 & 5.90 & 21.80 \\ 
  2021-12 & 1.20 & 1.20 & 3.70 & 4.50 & 4.40 & 6.00 & 2.70 & 28.90 \\ 
   \hline
\end{tabular}
}
\caption{Intersection in percentage of annual GFC tree loss data and our bi-annual deforestation product.}
\label{tab3}
\end{table}

\begin{figure}[ht]
 \centering
 \includegraphics[width=1\linewidth]{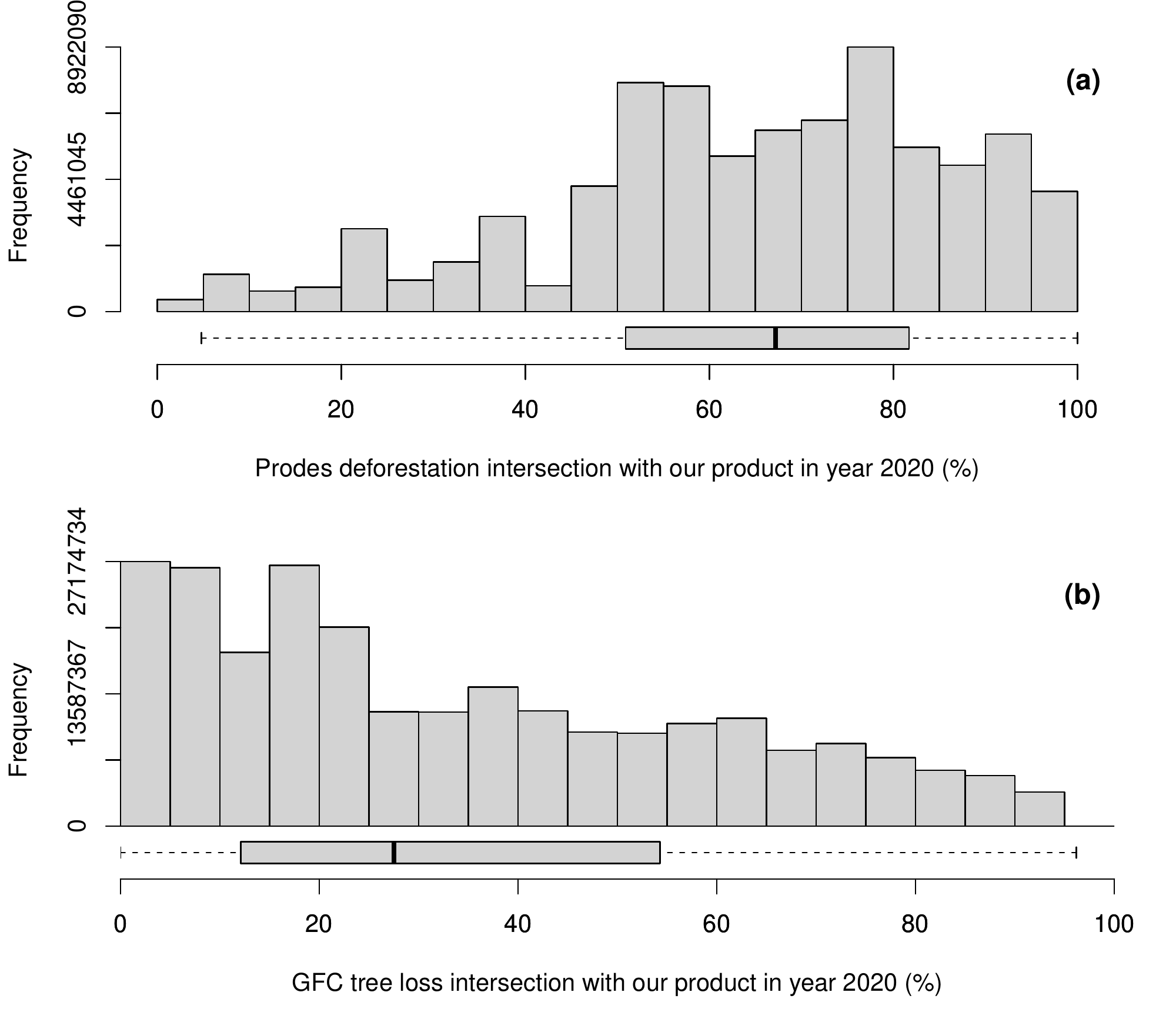}
  \caption{Histogram of intersection per Planet tiles of our product of deforestation with the Prodes data (a) and GFC tree cover loss (b) per quad for the year 2020. Each intersection value was weighed by the number of pixels with intersection observed in the tile, of Prodes or of GFC tree loss year data,  respectively.}
 \label{Fig4}
 \end{figure}

For the year 2020, the histogram of the percent of intersection per quad for our product and the Prodes deforestation data shown that our product share consistent information with the Prodes data, Fig. \ref{Fig4}a. The median intersection of Prodes with our product was  67.2\%, and 75\% of the pixels had more than 50.9\% of intersection  and 25\% had more than 81.7\%. Note that the Prodes data are manually made with data at 20 to 30 m spatial resolution and do not consider deforested area below 6.25 ha. Our product show less agreement with the GFC tree loss year data, Fig. \ref{Fig4}b. In 2020, only 25\% of our data had an intersection above 54\%. An important proportion of the GFC tree loss data of 2020 show almost no intersection with our product as shown by the median of intersection of 27.5\%. Still, for some quad the intersections can be above 80\%, Fig. \ref{Fig4}b.     

 \begin{figure}[ht]
 \centering\includegraphics[width=0.45\linewidth]{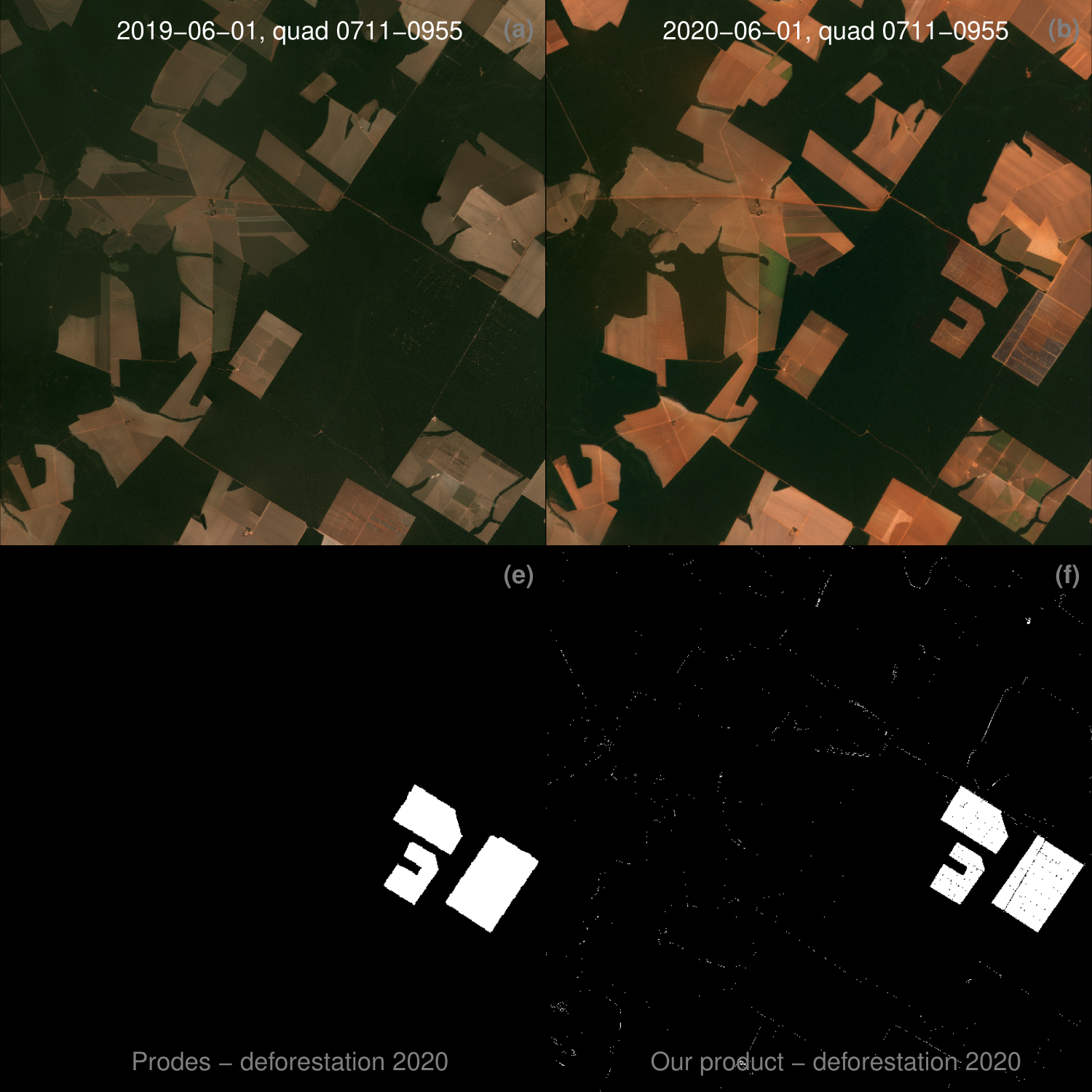}
 \centering\includegraphics[width=0.45\linewidth]{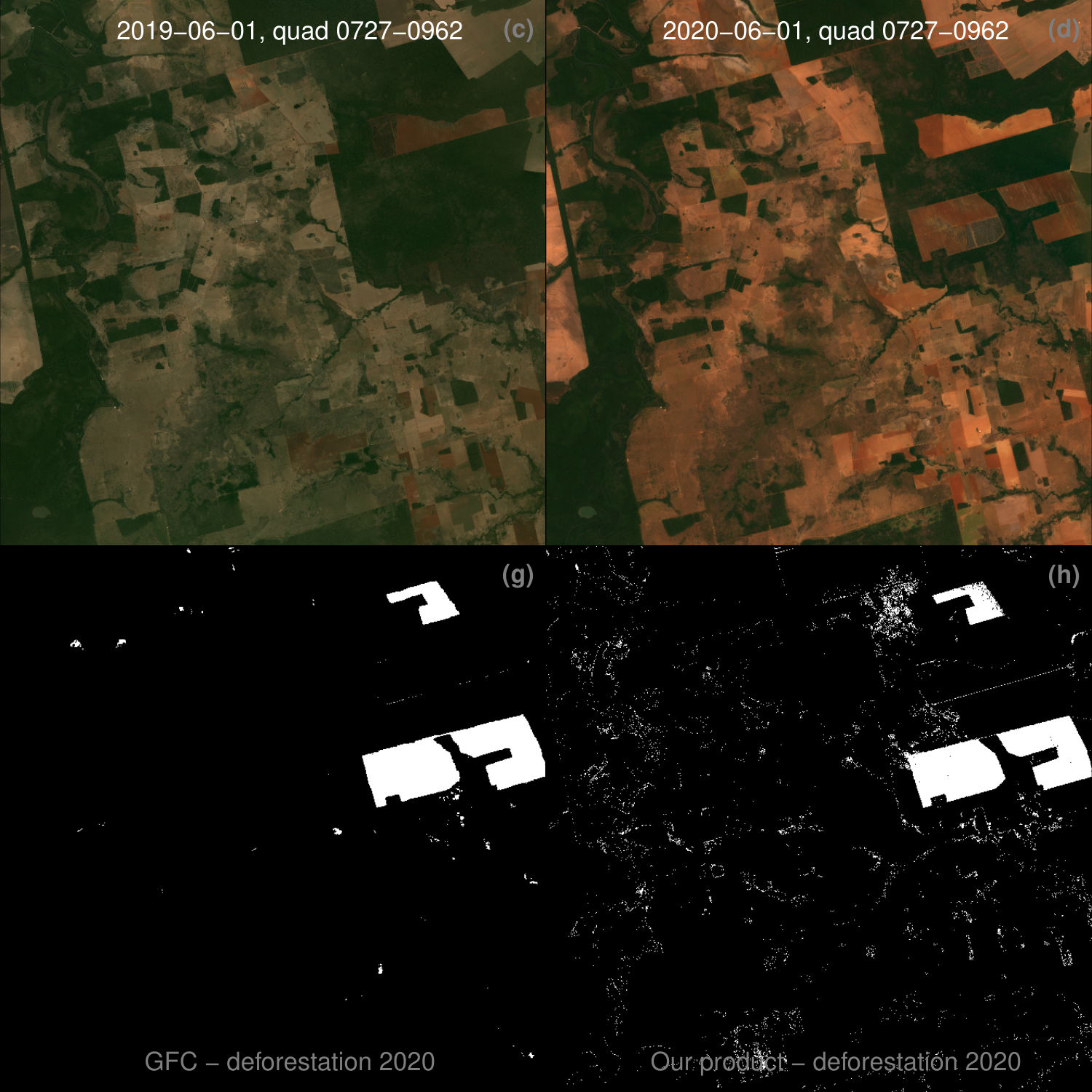}

  \caption{Example of good agreement, i.e. $>$ 90\% of intersection, of our product of deforestation with the Prodes data (Planet quad 0711-0955) and of our product and GFC tree loss year data (Planet quad 0727-0962) for the year 2020. Planet NICFI  quad is $\sim$ 19.5 $\times$ 19.5 km.}
 \label{Fig5}
 \end{figure}

For the tiles with similar results with our deforestation map, that is with more than 90\% of intersections between our deforestation mapping and the Prodes or GFC tree loss year data, Fig. \ref{Fig5}, several observations can be made. Between Prodes and our deforestation data, Fig. \ref{Fig5} a-b for the images and e-f for the deforestation masks, it can be observed that Prodes (as expected by the Prodes methodology) included non-forest pixels that are captured by our algorithm with Planet image at 5 m, for example the logging decks. It can also be observed that our algorithm detected deforested pixels on the borders of the forest patches and roads. It indicates that the pixel attribution has changed, however, it remains difficult to conclude if this is because of real change in tree cover or because of imprecise georeferencing of Planet images between the two dates. Between GFC tree loss year data and our product, Fig. \ref{Fig5} c-d for the images and g-h for the deforestation masks, it can be observed that GFC tree loss year data produced large patches and also some roads are detected, while our product detected a lot more changes in pixels. Inside deforestation patches, sometimes some tree remains, and our algorithm still detect them as forests. Overall Prodes and GFC tree loss year data give more homogeneous surfaces, and our product, with its finer spatial resolution gives a tree cover at the scale of the tree and small gap in the canopy can be marked as non-forest or few remaining trees inside a deforested patch can be still recognized as tree cover. Outside the forest the change of attribution of our product is difficult to interpret, as it can be due to deforestation or to imprecise Planet images georeferencing.

 \begin{figure}[ht]
 \centering\includegraphics[width=0.45\linewidth]{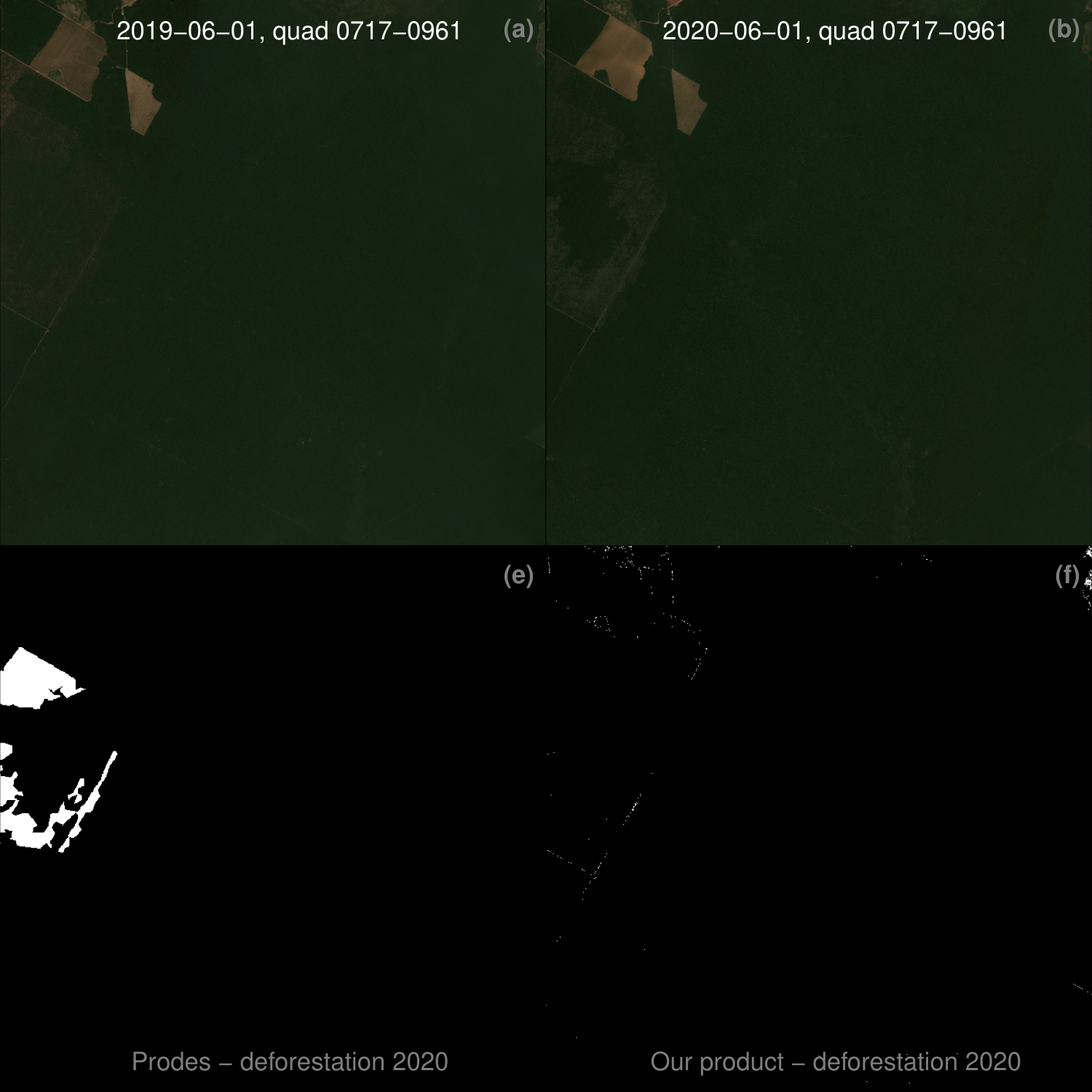}
 \centering\includegraphics[width=0.45\linewidth]{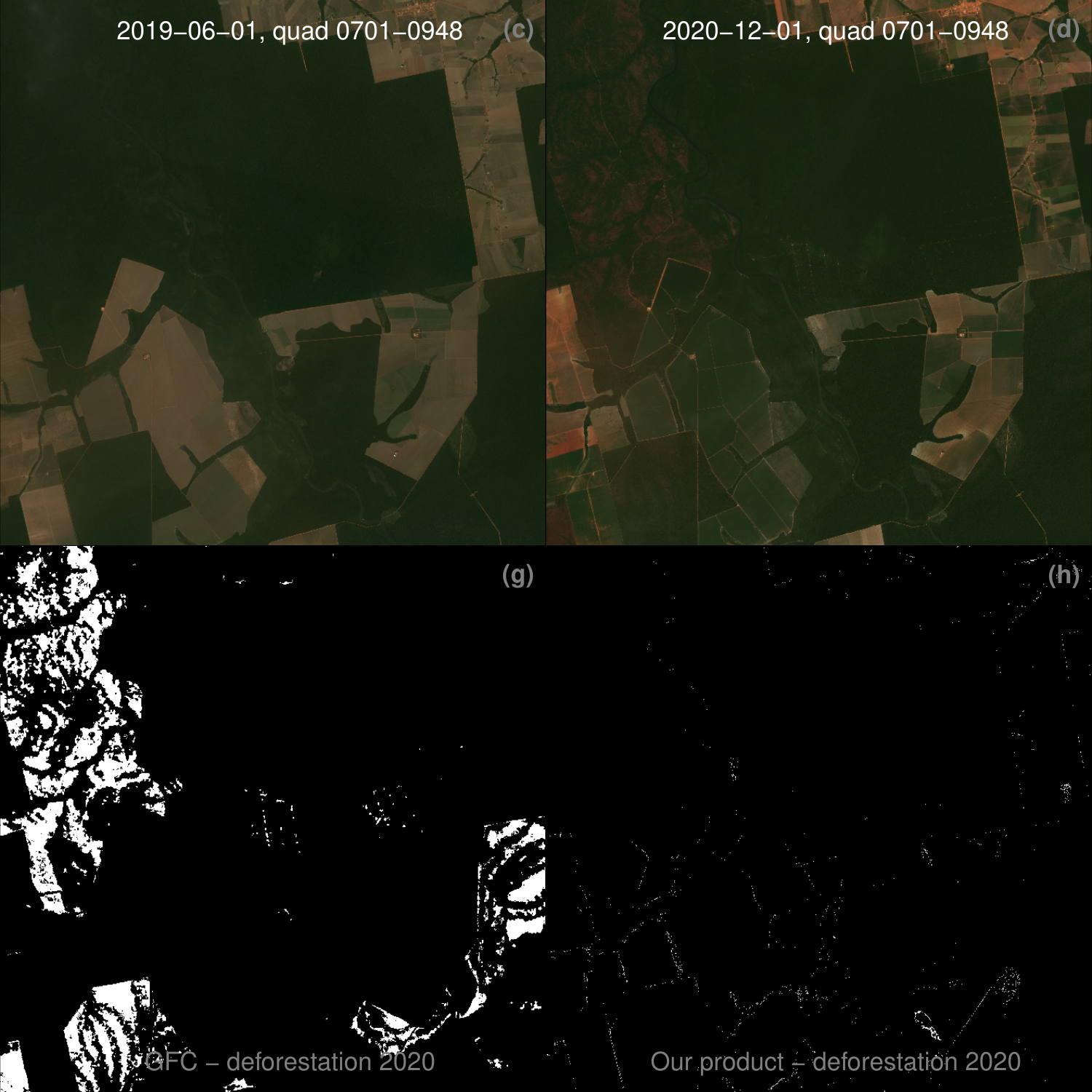}
  \caption{Example of poor agreement, i.e. $<$ 10\% of intersection, of our deforestation product and the Prodes data (Planet quad 0717-0961) and GFC tree loss year data (Planet quad 0701-09948) for the year 2020. Planet NICFI  quad is $\sim$ 19.5 $\times$ 19.5 km.}
 \label{Fig5.2}
 \end{figure}


For the tiles with bad agreement between our deforestation mapping and the Prodes or GFC tree loss year data, that is with less than 10\% of intersection, Fig. \ref{Fig5.2}, fire mapped as deforestation in Prodes and in GFC tree loss year data seems to be the culprit. While our algorithm mapped almost no change for the quad 0717-0961 in 2020, Fig. \ref{Fig5.2} a-b  and e-f, a large area of forest that have burnt is delineated as deforestation in Prodes, while it is still forest. The same observation is made for the GFC data for the region of the quad 0701-0948 in 2020, Fig. \ref{Fig5.2} c-d and g-h. In this tile, large fires have been mapped as deforestation, while the forest is still here. These fires included in the mapping of deforestation in Prodes or GFC tree loss year data leads to major discrepancy with our product.

 \begin{figure}[ht]
 \centering\includegraphics[width=0.45\linewidth]{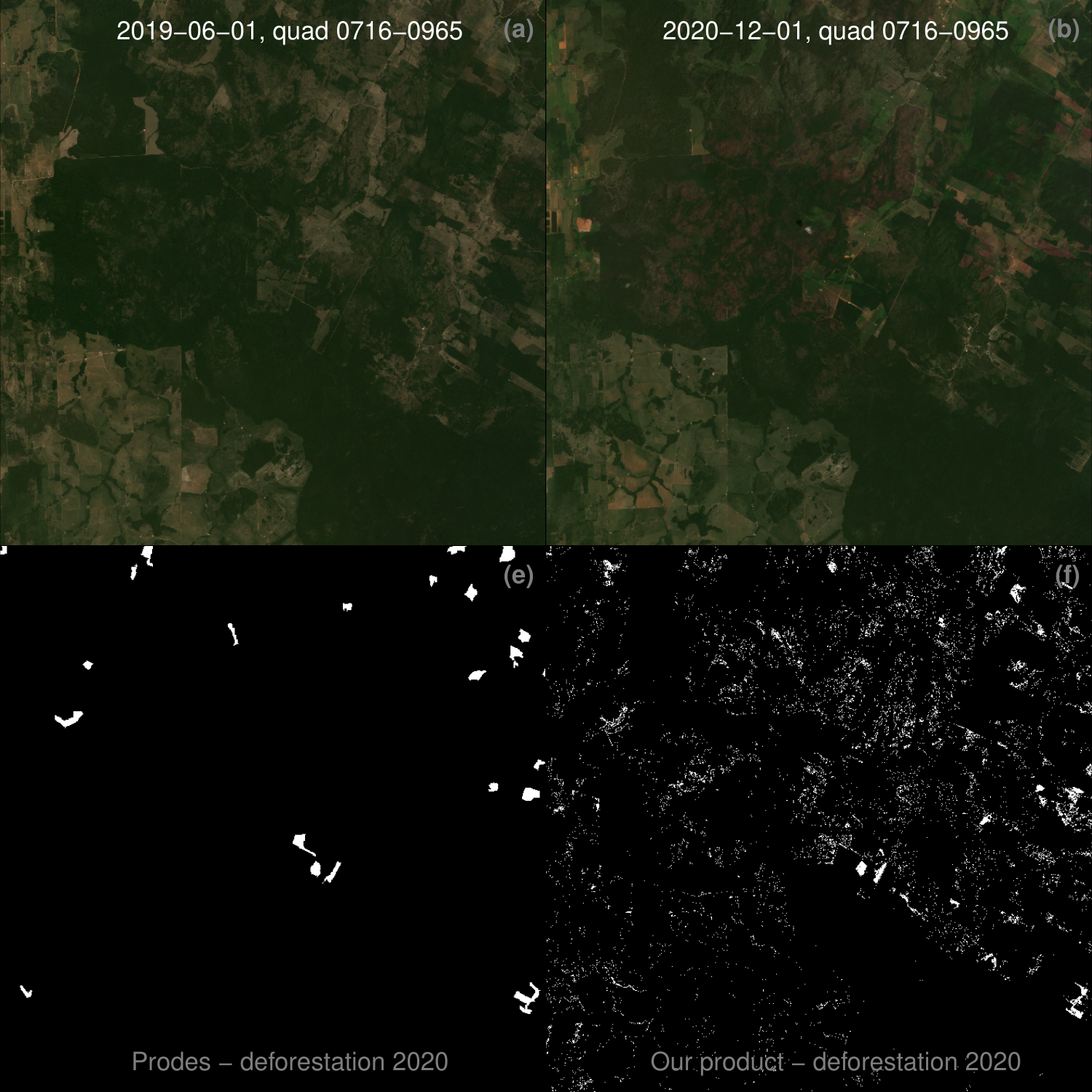}
 \centering\includegraphics[width=0.45\linewidth]{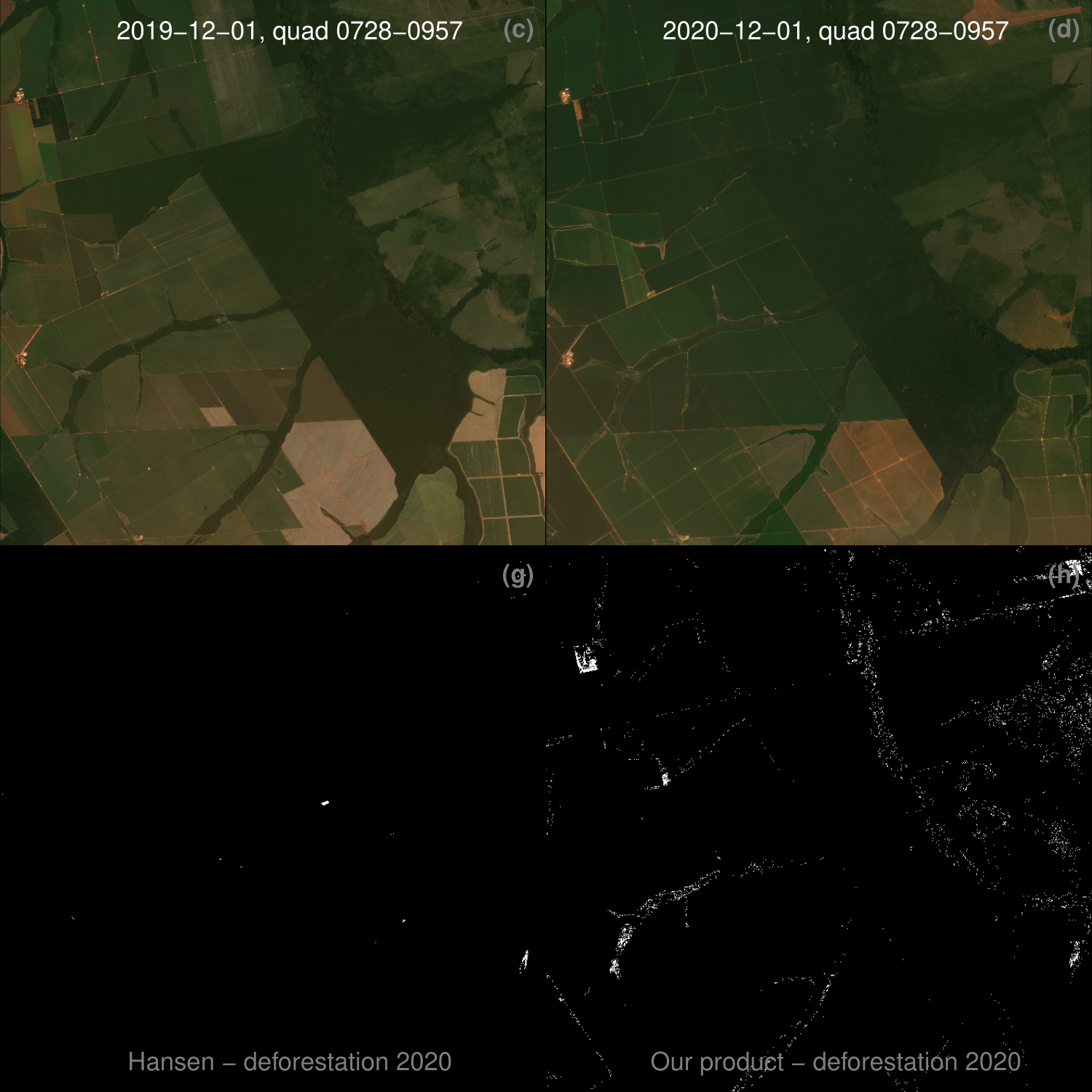}
  \caption{Worst cases of non - agreement, between Prodes data (Planet quad  0716 - 0965), Hansen tree cover loss (Planet quad 0728 - 0957)  and our product, i.e. $<$ 10\% of intersection,  for the year 2020. Planet NICFI quad is $\sim$ 19.5 $\times$ 19.5 km.}
 \label{Fig5.3}
 \end{figure}

For the tiles that shows the most extreme difference, that is our product find a lot more pixels  than Prodes or GFC tree loss year data ($>$100000 pixels), Fig \ref{Fig5.3}, it can be observed that the changes are mainly isolated pixels. It is also common around rivers and on the border of forests, where our algorithm can pick up really small changes as the spatial resolution is 5 m. However, it remains difficult to say if it is an anthropic or natural change or even a referencing problem. This also seems to happen for short forest stand that are not highly packed.

 \begin{figure}[ht]
 \centering
 \includegraphics[width=0.9\linewidth]{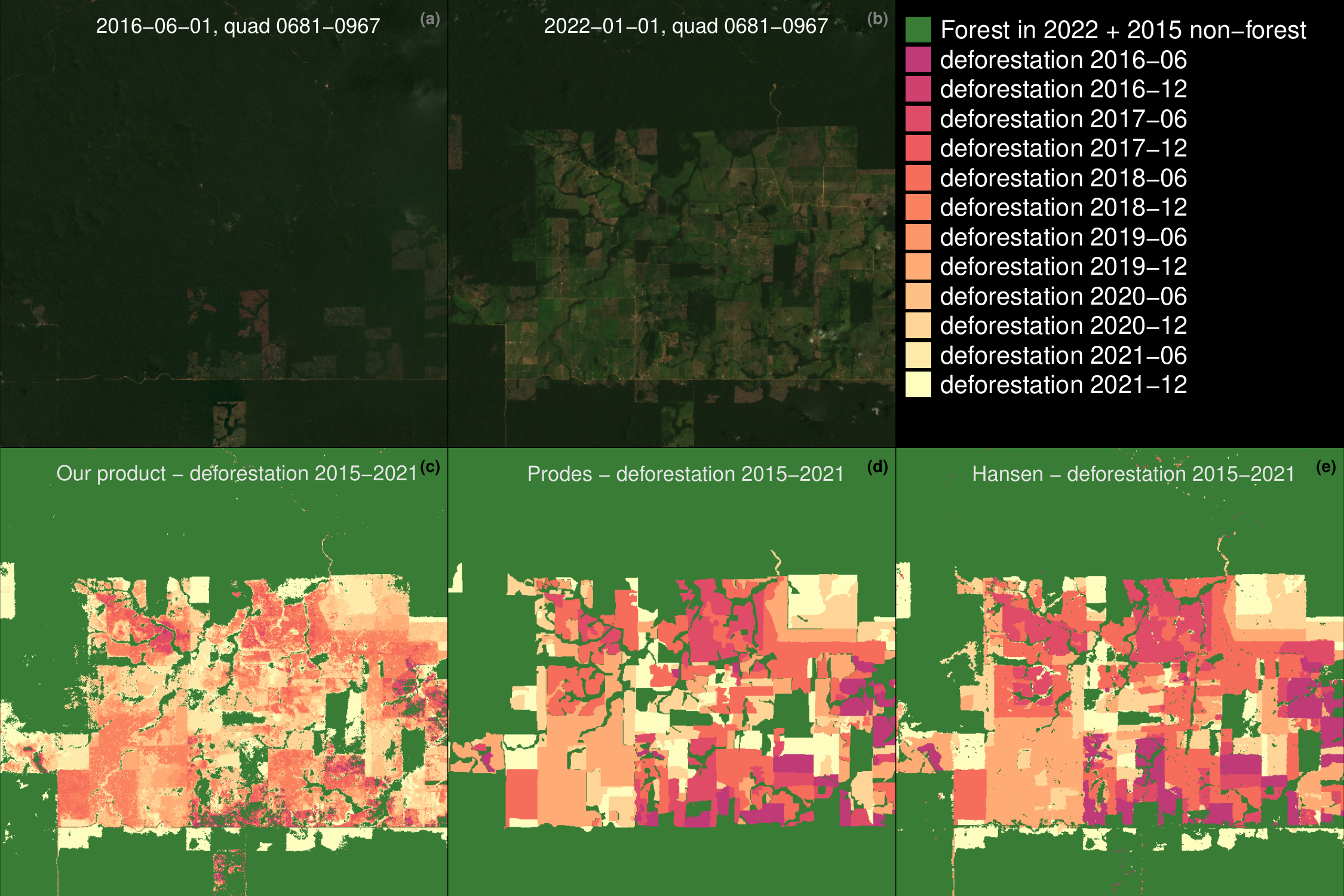}
  \caption{Comparison of our product of deforestation with the Prodes data (a) and GFC year of tree loss for the Planet quad 0681-0967 for the period from 2015-12-01 to 2021-12-01. Planet NICFI quad is $\sim$ 19.5 $\times$ 19.5 km.} 
 \label{Fig6}
 \end{figure}

Comparing the deforestation time series between our product, GFC tree loss year data and Prodes product, it can be observed that our product tends to delay the timing of deforestation for the first years of the time series, even if it detected well these areas spatially. This is due to the cloud filter that consider three dates around each date to filter for clouds. For example, for the year 2015, the three next data are 2016-06-01, 2016-12-01 and 2017-06-01 for the cloud filter. This issue is resolved after 2020-12-01, as the first date of the monthly time series is 2020-09-01 and will not appear for the future dates. For the last periods, our product is well aligned with Prodes and GFC tree loss year data tree loss year dataset. Our product appears noisier, because of the temporal resolution (6 months) and also the spatial resolution (5 m) while Prodes and GFC tree loss year data produce more homogeneous area. With our product is is possible to see the deforestation each 6 months. As our product does not use expert interpretation, the results are fast to produce, it just depends on the release of the image and the deforestation of the end of 2021 is included in our dataset, as for GFC tree loss year data but not in Prodes data (Prodes year end in August, our study was made in May). Our product missed some small regions, this is mainly due to remaining trees, as if some tree are remaining, our model still consider that a forest is present, while in 30 m resolution the change of signal between the two dates must be very high and easy to detect. Overall, our results seem to indicate that the time series of deforestation can be obtained with the Planet data at least when the Planet time series is monthly.

\section{Discussion}

\subsection{Mapping dense tropical tree cover}

In the Mato Grosso state - Brazil, the U-net network identified dense tropical forests with an overall accuracy and a F1-score above 98\% on the validation dataset, showing again the high capacity of deep learning to support vegetation mapping \cite{KATTENBORN202124} even in tropical environment \cite{Wagner2019, wagner2020m}. The 2015 tree cover represented 58.1 \% of the state. We found that 43.3 \% of Mato Grosso in 2021 is still forested with forest that was present in 2015. Note that our model does not account for regrowing forest. The high performance of the segmentation could be explained by the unique spectral values and textural information of the dense forests when considering the context, for example Fig. \ref{Fig5}-\ref{Fig6}. The k-textures model, that is used to make the training sample, was previously shown to segment these forests by textures without human supervision, which further confirm that they contain unique feature characteristics \cite{Wagner2022}. Here, the accuracy shown the robustness and generalization of the U-net model to such segmentation task. No heavy pre-processing was used, only a scaling to have 8 bits images and a border added with mirroring image to avoid border effect during the prediction. This minimal pre-processing eased the processing of large image database, such as the database of 85056 images used here. Planet images have high spectral variability due to lack of inter-sensor calibration between satellites, as well as the intra-annual variation of sunlight and atmospheric conditions present in the daily imagery that are combined to make the monthly or bi-annual composites  \cite{planet2021}. As only a light data augmentation was used, the generalisation of the model seems to confirm that natural variation of reflectance present in the satellite images is sufficient to improve the accuracy of the model and to help the model to generalize \cite{Wagner2021}. Our model was very robust to estimate non-forest as shown by the independent validation with LiDAR data, Fig. \ref{Fig2}a and made only few false positive errors ($\sim$5\%) on the forest cover, Fig. \ref{Fig2}b. We are less preoccupied by false positives than false negatives, that is, we don't want to miss any forested pixels. False positives can be easily corrected later with the temporal filter. For example, with a simple rule, i.e., the forest attribution cannot change from forest to non-forest and to forest again. Relatively few other observed errors were mainly sparse trees and with dark green background. As our algorithm look at the context, if there are still some trees, it can be classified as a forest, and as it does not look at the change between date, the model is unaware of the diminution of the number of trees and can miss the detection. This is a rare case and is more related to degradation that to deforestation. Predicting the tree cover at each date independently of the other dates is a major difference with models like GFC map \cite{Hansen850} where the authors looked at the difference between date to identify change. In future work, the deep learning model with Planet NICFI images could be adapted to work with the times series, such as works made to detect deforestation with Landsat \cite{matosak2022mapping,maretto2020sp}, and to account for more subtle change in tree cover. As the tree cover is initially self-segmented by the k-textures algorithm, an expert only define which classes found by the k-textures algorithm are representing tree cover, it is still a challenge to relate our tree cover meaning to the official forest definitions such as the FAO definition: 'Land spanning more than 0.5 hectares with trees higher than 5 meters and a canopy cover of more than 10 percent, or trees able to reach these thresholds in situ. It does not include land that is predominantly under agricultural or urban land use' \cite{FAO2020}. The U-net model needs a lot of data to train, and we still do not dispose of sufficient LiDAR dataset to try to get closer to the official definition. In the future, this could be done, by training a U-net including the height from LiDAR, for example.

 \subsection{Mapping of deforestation}

The tree cover maps at very high accuracy for all Planet dates enabled to reconstruct the history of deforestation for the region on the period 2015 - 2021 at a spatial resolution of $\sim$ 5 m and temporal resolution of 6 months. We observed that deforestation is still widespread in Mato Grosso and the deforestation front are actives, Fig . \ref{Fig3}. Our numbers show that deforestation is still alarmingly increasing is this state, with the tree cover change for the second semester of 2021 representing $\sim$2\% (19817.78 km$^2$) of the size of the state, Table \ref{tab1.2}. Several types of tree cover changes were observed, from small scale deforestation to large scale deforestation, roads, and also natural changes such as changes in river paths, Fig \ref{Fig9}. One of the major differences with GFC tree cover loss year and Prodes product \cite{Hansen850,INPE2021}, is that the entire tree cover is mapped at each date and not only the change in relation to a fixed previous year (2000 in the case of the GFC map) or previous dataset (deforested areas for Prodes) so we can compute the actual deforestation number. Our product is also bi-annual when other products are annual and in the next version, we will integrate a cloud mask specific to Planet NICFI image to have a monthly product. Large areas mapped by Prodes or GFC tree cover loss year that where not found by our product were mainly linked to forest fire or degradation due to fire mapped as deforestation, Fig. \ref{Fig5.2}.
We considered as tree cover pixelx that were classified as tree cover on half of the dates on a moving windows of $\pm$ three months. Consequently, it is possible that vegetation that is leafless during drought months start to be detected only with the monthly dataset (since 2020-09-01) and be classified as deforestation, which could increase the detection for the first years of the monthly dataset. This artifact should be resolved with a longer monthly time series. Furthermore, rivers roads and clearcut boundaries are also mapped with increased accuracy during the time series as it can take some time to see them clearly due to minor errors in the co-registration and the satellite view angles of images at different dates \cite{francini2020near}. For example, a river can appear as deforestation for the first dates of the series, until the larger path of the river is mapped. This artifact should also be also resolved with more monthly data. The monthly time series is still really new, and some more months would be needed to properly calibrate the deforestation map. Furthermore, our map also accounts for natural changes in tree cover, and not all changes that we observe are human made tree cover changes.  This can explain, in addition to the spatial resolution of 5 m, why our model maps up to 7.49 times more change in tree cover than the 30 m products, Table \ref{tab1.2}. In future work, we will also include regrowing forest, as there is a significant amount of regrowing forests and they are very interesting for carbon sequestration, with the potential of contributing ~5.5\% to Brazil’s 2030 net emissions reduction target \cite{silva2020b, heinrich2021, rosan2019e}. As for other tropical forest monitoring systems based on optical imagery, our model can be still limited on areas with high cloud cover and for fast detection of deforestation, for early warning systems, radar monitoring could be preferred as they are now efficient and used such as DETER-SAR in Brazil \cite{Doblas2022}. In future work, we will adapt the methodology to radar images to circumvent the cloud cover issues. Besides supporting forest monitoring and carbon assessment, our 5 m resolution tree cover map could also support forest fragmentation analysis for conservation as forest fragmentation is currently increasing in the Amazon \cite{montibeller2020,Vogt2009,Strassburg2016}.
Finally, our work demonstrates the potential of deep learning for mapping deforestation with Planet NICFI image.

\subsection{Advances in training sample production for land cover}
Here, we used an alternative to produce the large training sample needed to parameterize the U-net model, which is usually made by hand. The production of our training sample was made in a large part automatically with the use of self-segmentation by the k-textures algorithm \cite{Wagner2022}. The k-textures algorithm provides self-supervised segmentation of an image in a k number of texture classes and is fully deep learning based. The manual steps in this process consisted of the choice of number of classes returned by the k-textures model, to identify which classes represented forest, to perform small corrections on the mapped results, and to find additional samples of atmospheric conditions, such as forests below thin clouds which does not need segmentation (only adding in the sample 256$\times$256 images with thin cloud cover over forest). The training sample was therefore produced with minimal human supervision, that is, without the need of manually mapping samples of pixels that belonged to the forest class. This enables to make the training sample very quickly, in a couple of days, and which could explain the high accuracy of segmentation. Furthermore, the k-textures already use several features and multiple levels of abstraction, to separate each class, and a human cannot be as consistent and fast \cite{BRODRICK2019734}. Drawing polygons manually cannot be as accurate as the model, as the manual delineation of pixels cannot follow the pixels in the borders of forests accurately, for example. As a result, the production of the sample was faster and the training sample more accurate. More generally, this approach could ease the production of training sample for classes that have a different texture from other land-covers.


\section{Conclusions}
In this work, we showed the great potential of the popular CNN semantic segmentation architecture U-net \cite{Ronneberger2015} to map tropical tree cover and deforestation with Planet NICFI images. The U-net accurately segmented dense tree cover (F1-score $>$ 0.98) with minimal preprocessing and even with the high reflectance variability of the  Planet NICFI images.  The mapping of tree cover at a 6-month scale enables to map of deforestation further and brings information on forest ecosystem change at an unprecedented time scale. In addition, our method detects small patches of deforestation, which allows us to locate areas in an early stage, helping institutions responsible for fiscalization and law enforcement to avoid illegal deforestation. The model will be further improved to work with images from in different tropical forest regions, more solar/view angles and atmospheric conditions by increasing the training sample and scaled on the cloud to provide global biannual tree cover and deforestation data for the entire  Planet NICFI dataset.

\section{Authors contributions}
Conceptualization, F.H.W. and S.S.; methodology, F.H.W., R.D., C.H.L.S.J, G.C., A.L.R., M.C.M.H.; software, F.H.W. and R.D.; validation, F.H.W., R.D. and J.P.H.B.O; formal analysis, F.H.W.; investigation, F.H.W. R.D. S.S.; resources, S.S.; data curation, F.H.W.; writing---original draft preparation, F.H.W., R.D., C.H.L.S.J, and S.S.; writing---review and editing, F.H.W., R.D., C.H.L.S.J and S.S; visualization, F.H.W.; supervision, S.S.; project administration, S.S.; funding acquisition, S.S. All authors have read and agreed to the published version of the manuscript.

\section{funding} This research received no external funding 

\section{Data availability}The TensorFlow 2 U-net model with trained weights is available on Zenodo \url{https://doi.org/10.5281/zenodo.7324753}. Planet NICFI data are freely available trough the Planet Mosaics API \url{https://developers.planet.com/docs/basemaps/reference/##tag/Basemaps-and-Mosaics}. 

\section{acknowledgments}
We thank the team from the EBA project for supporting the use of the airborne LiDAR datasets. Part of this work was carried out at the Jet Propulsion Laboratory, California Institute of Technology, under a contract with the National Aeronautics and Space Administration (NASA).

\section{Conflicts of interest}
The authors declare no conflict of interest. The funders had no role in the design of the study; in the collection, analyses, or interpretation of data; in the writing of the manuscript; or in the decision to publish the~results.

\bibliographystyle{unsrtnat}
\bibliography{references}  






\end{document}